\documentclass[twocolumn,preprintnumbers,amsmath,amssymb]{revtex4}
\usepackage{dcolumn}
\usepackage{bm}
\usepackage{graphicx}
\usepackage{amsmath}
\begin{document}

\title{Stability and collisions of quantum droplets in $\mathcal{PT}$-symmetric dual-core couplers}
\author{Zheng Zhou$^{1,2}$, Bo Zhu$^{3}$, Haibin Wang$^{1}$, and Honghua Zhong$^{4}$\footnote{corresponding author, E-mail: hhzhong115@163.com.}}
\affiliation{$^{1}$Department of Physics, Hunan Institute of Technology, Hengyang 421002, China\\
$^{2}$College of Physics and Electronic Engineering, Hengyang Normal University, Hengyang 421002, China\\
$^{3}$School of Physics and Astronomy, Sun Yat-Sen University (Zhuhai Campus), Zhuhai 519082, China\\
$^{4}$Institute of Mathematics and Physics, Central South University of Forestry and Technology, Changsha 410004, China}

\begin{abstract}
We study the effect of the interplay between parity-time ($\mathcal{PT}$) symmetry and optical lattice (OL) potential on dynamics of quantum droplets (QDs) forming in a binary bosonic condensate trapped in a dual-core system. It is found that the stability of symmetric QDs in such non-Hermitian system depends critically on the competition of gain and loss $\gamma$, inter-core coupling $\kappa$, and OL potential. In the absence of OL potential, the $\mathcal{PT}$-symmetric QDs are unstable against symmetry-breaking perturbations with the increase of the total condensate norm $N$, and they retrieve the stability at larger $N$, in the weakly-coupled regime. As expected, the stable region of the $\mathcal{PT}$-symmetric QDs shrinks when $\gamma$ increases, i.e., the $\mathcal{PT}$ symmetry is prone to break the stability of QDs. There is a critical value of $\kappa$ beyond which the $\mathcal{PT}$-symmetric QDs are entirely stable in the unbroken $\mathcal{PT}$-symmetric phase. In the presence of OL potential, the $\mathcal{PT}$-symmetric on-site QDs are still stable for relatively small and large values of $N$. Nevertheless, it is demonstrated that the OL potential can assist stabilization of $\mathcal{PT}$-symmetric on-site QDs for some moderate values of $N$. On the other hand, it is worth noting that the relatively small $\mathcal{PT}$-symmetric off-site QDs are unstable, and only the relatively large ones are stable. Furthermore, collisions between stable $\mathcal{PT}$-symmetric QDs are considered too. It is revealed that the slowly moving $\mathcal{PT}$-symmetric QDs tend to merge into breathers, while the fast-moving ones display quasi-elastic collision and suffer fragmentation for small and large values of $N$, respectively.
\end{abstract}

\maketitle

\section{Introduction}
In the last few years, self-bound quantum droplets (QDs) in ultracold atoms have attracted increasing attention from both experimental and theoretical research \cite{Baillie255302, Wachtler061603, Baillie021602, Barbut160402, Ferioli090401, Kartashov193902, Tengstrand160405, Errico033155, Chiquillo051601, Mishra073402, Ferioli013269, Oldziejewski090401, Zhang133901}. In the pioneering theoretical work \cite{Petrov155302}, the three-dimensional (3D) QDs were predicted as stable soliton-like states in binary Bose-Einstein condensates (BECs) due to the balance between the attractive mean-field interaction (the inter-component attraction being slightly stronger than the intra-component repulsion) and the repulsive Lee-Huang-Yang (LHY) correction originating from quantum fluctuation \cite{Lee1135}, on the basis of a system of coupled mean-field Gross-Pitaevskii equations. While intriguing breakthrough in the field of the dipolar BECs \cite{Griesmaier160401, Beaufils061601}, the first experimental observations of QDs have been realized in dipolar Bose atoms \cite{Kadau194, Barbut215301, Schmitt259, Chomaz041039, Saito053001}, by exploiting the competition of long-range dipole-dipole attraction and LHY repulsion. Very recently, tuning contact interactions through the Feshback resonance technique, QDs have been also observed in a binary condensate of two different atomic states of $^{39}$K atoms, both in the presence of an external potential \cite{Cheiney135301, Cabrera301} and in free space \cite{Semeghini235301}, which exactly follow the original theoretic predictions \cite{Petrov155302}.

It is worth noting that the QDs are more ubiquitous and remarkable in a lower dimensionality. The reduction of the dimension from 3D to 2D drastically changes the form of the LHY term, replacing the quartic form by the cubic terms multiplied by a logarithmic factor \cite{Petrov100401}. It was recently predicted that 2D QDs with embedded vorticity $S$ may be stable up to $S=5$ \cite{Li113043}. In the 1D geometry, the sign of LHY term is changed from repulsive in higher dimensions (2D and 3D) into attractive in 1D. Accordingly, the 1D stable QDs can be formed by the competition of the effective cubic mean-field repulsion and the quadratic LHY-induced attraction. In particular, the stability and collisions of 1D Gaussian-shaped and flat-top QDs in free space were recently studied in the case of relatively small and large numbers of atoms, respectively \cite{Astrakharchik013631}. Furthermore, due to the availability of optical lattices (OLs) for experiments with BEC, the next natural step is the consideration of dynamics of 1D QDs trapped in OL potential \cite{Zhou104881}. Besides, the spontaneous symmetry breaking of 1D QDs in a dual-core trap has also been considered \cite{Liu053602}.

On the other hand, the interaction with environment plays an important role in studies of ultracold atoms, which renders the system non-Hermitian. A special class of non-Hermitian systems that gained much interest, since the seminal work by Bender and Boettcher \cite{Bender5243}, are parity-time ($\mathcal{PT}$)-symmetric ones, where $\mathcal{P}$ denotes the parity reflection operator $\hat{x}\rightarrow-\hat{x}$, $\hat{p}\rightarrow-\hat{p}$ and $\mathcal{T}$ the time reversal operator $\hat{x}\rightarrow-\hat{x}$, $\hat{p}\rightarrow-\hat{p}$, $i\rightarrow-i$, $t\rightarrow-t$. The most characteristic property of $\mathcal{PT}$-symmetric systems is the existence of an entirely real eigenvalue spectrum within a certain parameter regime \cite{Bender947}. Due to the equivalence between the Schr\"{o}dinger equation and the equations describing the propagation of light \cite{Longhi243}, optical systems with complex refractive indices \cite{Makris103904, Musslimani030402} are widely used to study $\mathcal{PT}$ symmetry in non-Herimitian system. The first experimental realization of $\mathcal{PT}$ symmetry is in optical systems \cite{Guo093902, Ruter192}. By contrast, a promising candidate for exploring the dynamics of a genuine quantum system with balanced gain and loss is a BEC in a double-well potential \cite{Klaiman080402}. The gain and loss is realized by adding atoms in one well and coherently removing particles in the other. Nevertheless, the coherent incoupling and outcoupling of particles is a very non-trivial task. In the last few years, the experimental realization of $\mathcal{PT}$ symmetry in BECs has been attracting increasing attention. It was proposed that the use of bounded and unbounded states to provide such a coherent in- and out-coupling of particles \cite{Single042123}. Besides, another double-well system was suggested as a particle reservoir for the implementation of $\mathcal{PT}$ symmetry \cite{Gutohrlein335302}. Furthermore, Kreibich $\emph{et al.}$ developed an experimental scheme to realize a $\mathcal{PT}$-symmetric two-mode system based on time-dependent optical lattice, by embedding this system in a larger multiwell system where the additional wells are considered as reservoir wells \cite{Kreibich051601, Kreibich033630, Kreibich023624}. However, these methods are difficult to realize experimentally because the experimental setup is quite demanding and currently hardly realizable. In more recent work, the realization of $\mathcal{PT}$-symmetric and $\mathcal{PT}$-symmetry-broken states in BECs of $^{87}$Rb atoms with a time-independent optical lattice was also presented \cite{Kogel063610}. Particularly, the $\mathcal{PT}$-symmetry-breaking transition was successfully observed in a gas of two-component noninteracting $^{87}$Li atoms in a passive way, i.e., there is only loss in the experiment. In the past decade, $\mathcal{PT}$-symmetric systems have been the subject in optics \cite{Driben4323, Peng394}, microwave cavities \cite{Bittner024101, Huai043803}, electronics \cite{Bender234101, Bender040101}, and ultracold atoms \cite{Li855, Cartarius013612, Dast124, Haag023601, Fortanier063608, Dast033617, Dast053601, Dast023625, Lunt023614, Haag033607, Zhou043412}, and so on. But up to date, the study of dynamics of QDs in $\mathcal{PT}$-symmetric systems is still lacking.

In this work, we aim to investigate the effect of the interplay between $\mathcal{PT}$ symmetry and OL potential on stability and collision dynamics of QDs trapped in the effectively 1D dual-core couplers. In Sec. \uppercase\expandafter{\romannumeral 2}, we introduce the model, and some analytical results for $\mathcal{PT}$-symmetric QDs are presented. Besides, the spontaneous $\mathcal{PT}$-symmetry-breaking phase transition is depicted. In Sec. \uppercase\expandafter{\romannumeral 3}, basic numerical results about stability analysis for $\mathcal{PT}$-symmetric QDs both in the absence of OL potential and in the presence of OL potential are reported. Collisions of $\mathcal{PT}$-symmetric QDs are addressed in Sec. \uppercase\expandafter{\romannumeral 4}. The paper is concluded by Sec. \uppercase\expandafter{\romannumeral 5}.

\section{The model}
We consider QDs forming in the binary condensate with mutually symmetric spinor components trapped in the 1D dual-core cigar-shaped potential with balanced gain and loss. The dynamics of QDs in such open dual-core setting is described by the system of linearly-coupled Gross-Pitaevskii equations for the wave functions in two cores, $\psi_1$ and $\psi_2$, including the cubic self-repulsion and the LHY-induced quadratic self-attraction in the scaled form, by using dimensionless units ($\hbar=m=1$) \cite{Petrov100401, Astrakharchik013631, Liu053602, Kogel063610, Li013604}:
\begin{eqnarray}\label{GPEs}
i\partial_t\psi_1=-\frac{1}{2}\partial_{xx}\psi_1+g|\psi_1|^2\psi_1-|\psi_1|\psi_1
+i\gamma\psi_1-\kappa\psi_2,\nonumber \\
i\partial_t\psi_2=-\frac{1}{2}\partial_{xx}\psi_2+g|\psi_2|^2\psi_2-|\psi_2|\psi_2
-i\gamma\psi_2-\kappa\psi_1,
\end{eqnarray}
where $g>0$ is the strength of the cubic self-repulsion, $\gamma$ is the gain and loss coefficient, and $\kappa>0$ is the rate of hopping between the two cores. By means of additional rescaling, we fix the strength of the cubic self-repulsion $g\equiv1$. Thus, the model is controlled by two irreducible parameters, $\gamma$ and $\kappa$. It is relevant to stress that, defining the parity operator as $\mathcal{P}$, which interchanges the two cores labeled by $1$ and $2$, and the time operator as $\mathcal{T}$: $i\rightarrow -i$, $t\rightarrow -t$, which reverses the time, the Hamiltonian for the system (1) is $\mathcal{PT}$ symmetric which fulfills $[\hat{H}, \mathcal{PT}]=0$. Previously, 1D and 2D $\mathcal{PT}$-symmetric dual-core couplers with cubic self-attraction and quintic repulsion in each core were introduced in optics \cite{Burlak113103, Burlak062904}.

The total norm of the system governed by Eq. (\ref{GPEs}), which is proportional to the number of atoms in the condensate, is
\begin{equation}\label{totalnorm}
N=\int_{-\infty}^{+\infty}(|\psi_1|^2+|\psi_2|^2)dx=N_1+N_2.
\end{equation}
Combining Eq. (\ref{totalnorm}) with Eq. (\ref{GPEs}), we can obtain the balance condition of the total norm
\begin{equation}\label{balance}
\frac{dN}{dt}=2\gamma(N_1-N_2),
\end{equation}
which demonstrates that only symmetric QDs, i.e., $N_1=N_2$, may represent stationary modes. The $\mathcal{PT}$-symmetric system cannot support stable asymmetric QDs as the balance between the gain and loss is impossible for them. The spontaneous symmetry breaking of QDs associating with bifurcation loop is not expected to occur in the $\mathcal{PT}$-symmetric system. Thus, a drastic difference of QDs in the  $\mathcal{PT}$-symmetric dual-core system from QDs in its conservative counterpart is that unstable symmetric QDs are not replaced by stable asymmetric QDs beyond the symmetry-breaking boundary \cite{Liu053602}.

\begin{figure}[htp] \center
\includegraphics[width=3.5in]{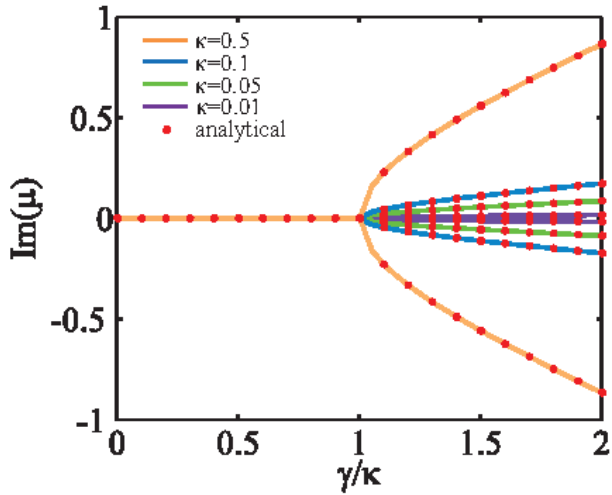}
\caption{\scriptsize{(Color online) The imaginary part of eigenvalues as a function of the ratio of the gain-loss parameter $\gamma$ and the inter-core coupling strength $\kappa$ for different values of $\kappa$, $\kappa=0.01$, $0.05$, $0.1$, and $0.5$, severally. The circular points indicate the analytical
results and the curves represent the numerical ones. Obviously, $\gamma=\kappa$ is the exceptional point in the $\mathcal{PT}$-symmetric dual-core trap.}}
\end{figure}

It is well known that the exact analytical solutions can give a deeper understanding than direct numerical simulations \cite{Zakrzewski3748, Dunlap3625}. To do so, we make the wave function transformation, provided that $\gamma/\kappa\leqslant 1$
\begin{equation}\label{transformation}
\psi_2=\left(i\frac{\gamma}{\kappa}+\sqrt{1-(\frac{\gamma}{\kappa}})^2\right)\psi_1
=\exp(i\sqrt{\kappa^2-\gamma^2}t)\psi,
\end{equation}
gives rise to the solvable standard nonlinear Schr\"{o}dinger equation with the cubic-quadratic nonlinearity
\begin{equation}\label{standard}
i\psi_t+\frac{1}{2}\psi_{xx}-|\psi|^2\psi+|\psi|\psi=0.
\end{equation}
Stationary symmetric QDs with a phase shift in the $\mathcal{PT}$-symmetric dual-core system can be found in the known form \cite{Astrakharchik013631}
\begin{equation}\label{exactsolution}
\psi(x)=\frac{-3(\mu+\kappa)\exp(-i\mu t)}{1+\sqrt{1+\frac{9}{2}(\mu+\kappa)}\cosh\left(\sqrt{-2(\mu+\kappa)}x\right)},
\end{equation}
where the chemical potential takes values in the interval of $-2/9<\mu+\kappa<0$. It is worth noting that, in the limit of $(\mu+\kappa)\rightarrow -2/9$, the symmetric QDs (\ref{exactsolution}) take an flat-top shape with a nearly constant wave function, $\psi=2/3$. The flat-top wave form is bounded by two fronts interpolating between zero and the constant wave function.

Note that the $\mathcal{PT}$-symmetric systems undergo phase transition as the gain-loss parameter crosses a certain threshold. Below this threshold, all eigenvalues are real (unbroken $\mathcal{PT}$-symmetric phase), but above this threshold, complex eigenvalues appear (broken $\mathcal{PT}$-symmetric phase). QDs amplify exponentially during time evolution in the broken $\mathcal{PT}$-symmetric phase, and any QDs would also be unstable to perturbations. Thus, We first consider the critical point for the phase transition between real and complex eigenvalues in the $\mathcal{PT}$-symmetric dual-core trap. To do so, we look for stationary QD solutions of the form
\begin{eqnarray}\label{stationarysolution}
\psi_1=\sum\limits_{n,k}\phi_{n,k}^A(x)\exp(-i\mu_{n,k}^A t),\nonumber\\
\psi_2=\sum\limits_{n,k}\phi_{n,k}^B(x)\exp(-i\mu_{n,k}^B t),
\end{eqnarray}
where $\phi_{n,k}^A(x)$ and $\phi_{n,k}^B(x)$ are the $n$-band Bloch wave function with quasi-momentum $k$, and $\mu_{n,k}^A$ and $\mu_{n,k}^B$ are the corresponding eigenvalues (also called chemical potential). The Bloch wave function can be defined as superposition of plane waves with different values of $k$
\begin{eqnarray}\label{Blochsolution}
\phi_{n,k}^A(x)=\sum\limits_{m}c_{n,k+Fm}^A\exp[-i(k+Fm)x],\nonumber\\
\phi_{n,k}^B(x)=\sum\limits_{m}c_{n,k+Fm}^B\exp[-i(k+Fm)x],
\end{eqnarray}
where $F$ is determined by lattice constant. Substituting Eqs. (\ref{stationarysolution}) and (\ref{Blochsolution}) into Eq. (\ref{GPEs}), we have
\begin{eqnarray}\label{modify}
&&\sum\limits_{m'}\left(
\begin{array}{clr}-\mu_{n,k}^A&~~~0
\\0
&-\mu_{n,k}^B
\end{array}
\right)
\left(
\begin{array}{clr}
c_{n,k+Fm'}^A \\c_{n,k+Fm'}^B
\end{array}\right)=
\sum\limits_{m}\left(
\begin{array}{clr}\hat{L}_1&~0
\\0
&\hat{L}_2
\end{array}
\right)\nonumber\\
&&\times\left(
\begin{array}{clr}
c_{n,k+Fm}^A \\c_{n,k+Fm}^B
\end{array}\right)+\sum\limits_{m}\left(
\begin{array}{clr}-i\gamma&k
\\k
&i\gamma
\end{array}
\right)
\left(
\begin{array}{clr}
c_{n,k+Fm}^A \\c_{n,k+Fm}^B
\end{array}\right),
\end{eqnarray}
with operators
\begin{eqnarray}\label{operators}
\hat{L}_1=-\frac{(k+Fm)^2}{2}-g\mid c_{n,k+Fm}^A\mid^2+\mid c_{n,k+Fm}^A\mid,\nonumber\\
\hat{L}_1=-\frac{(k+Fm)^2}{2}-g\mid c_{n,k+Fm}^B\mid^2+\mid c_{n,k+Fm}^B\mid,
\end{eqnarray}
we define\label{Hamiltonian}
\begin{eqnarray}
\hat{H}_1=\left(
\begin{array}{clr}\hat{L}_1&~0
\\0&\hat{L}_2
\end{array}
\right),
\hat{H}_2=\left(
\begin{array}{clr}-i\gamma&k
\\k&i\gamma
\end{array}
\right).
\end{eqnarray}
It is worth noting that the imaginary part of eigenvalues is affected only by $\hat{H}_2$. By diagonalizing $\hat{H}_2$, the two eigenvalues and eigenvectors are easily found to be $\mu_{\pm}=\pm\sqrt{\kappa^2-\gamma^2}$ and $\psi_{\pm}=(-i\gamma\pm\sqrt{\kappa^2-\gamma^2},\kappa)^T$ (the superscript $T$ represents transposition). Obviously, the two eigenvalues switch from real to complex values when $\gamma=\kappa$. Such a point ($\gamma=\kappa$) where both eigenvalues and eigenvectors coalesce is often referred to as an exceptional point (EP) \cite{Kato1966, Heiss2455}. EPs have turned out to be at the origin of many counterintuitive phenomena appearing in non-Hermitian systems that experience gain or loss \cite{Cao61, Heiss444016}. Very recently, the topological nature \cite{Lee133903, Leykam040401, Shen146402} and dynamical phenomena \cite{Hassan093002, Doppler76, Xu80} around the EPs have also been explored. The spontaneous $\mathcal{PT}$-symmetry-breaking transition that occurs at the EP can also be understood by examining the corresponding linear problem of Eq. (\ref{GPEs}), i.e., $i\partial_t\psi_{1,2}=-\frac{1}{2}\partial_{xx}\psi_{1,2}\pm i\gamma\psi_{1,2}-\kappa\psi_{2,1}$, following the procedure in Refs. \cite{Makris103904, Musslimani030402, Makris063807, Nixon023822}. We numerically show the imaginary part of eigenvalues as a function of the gain-loss parameter $\gamma$ for different values of the inter-core coupling strength $\kappa$, such as $\kappa=0.01$, $0.05$, $0.1$, and $0.5$, as shown in Fig. 1, where the curves represent the numerical results and the circular points correspond to the analytical results. The analytical results are in agreement with the numerical ones. In the broken $\mathcal{PT}$-symmetric phase ($\gamma/\kappa>1$), any QDs are unstable to perturbations, and there exists an overall destabilizing effect on propagation. In the following, we investigate the dynamics of symmetric QDs in the unbroken $\mathcal{PT}$-symmetric phase ($\gamma/\kappa<1$).

\begin{figure}[htp] \center
\includegraphics[width=3.5in]{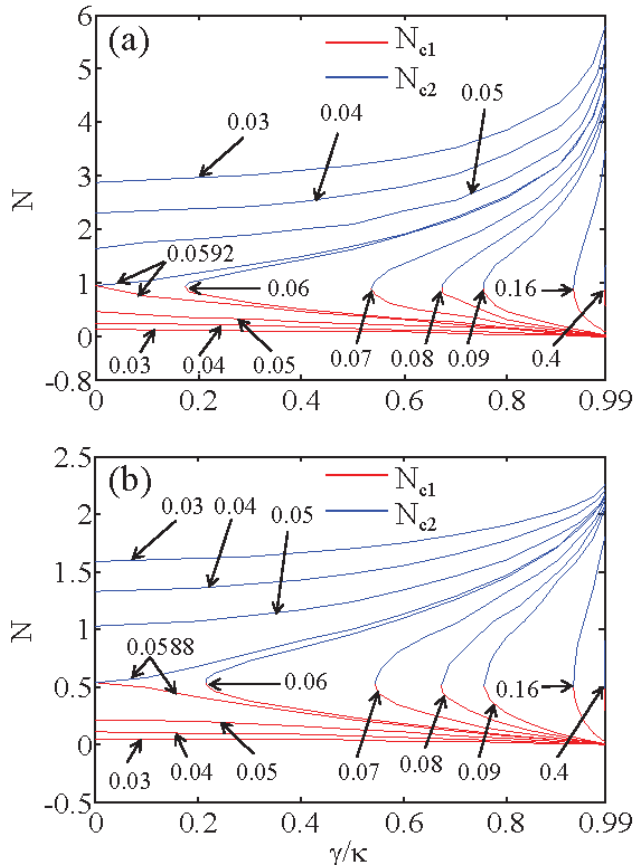}
\caption{\scriptsize{(Color online) Stability border for the $\mathcal{PT}$-symmetric QDs at different values of $\kappa$, which are indicated by arrows. The $\mathcal{PT}$-symmetric QDs with total norm $N$ are unstable in the interval of $N_{c1}<N<N_{c2}$ (branched $N_{c1}$ and $N_{c2}$ correspond to red and blue lines, respectively), and are stable at $N<N_{c1}$ and $N>N_{c2}$. (a) $\mathcal{PT}$-symmetric QDs in the absence of OL potential. For $\kappa<\kappa_c\approx 0.0592$, branches $N_{c1}$ and $N_{c2}$ are completely separated. For $\kappa_c<\kappa<\kappa_{max}\approx 0.4$, the branches $N_{c1}$ and $N_{c2}$ partly merge, and all the $\mathcal{PT}$-symmetric QDs are stable at $\gamma/\kappa<(\gamma/\kappa)_c$ corresponding to the merger point of the $N_{c1}$ and $N_{c2}$ branches. For $\kappa>\kappa_{max}$, the $\mathcal{PT}$-symmetric QDs become entirely stable in the whole interval of $0\leqslant\gamma/\kappa\leqslant 0.99$. (b) $\mathcal{PT}$-symmetric on-site QDs in the presence of OL potential $V=V_0\cos^2(\frac{\pi}{D}x+\theta)$, and the parameters are set as $V_0=0.3$, $D=8$, and $\theta=\pi/2$. There exist still two stability areas, nevertheless, the two threshold values $N_{c1}$ and $N_{c2}$ become smaller by comparing Fig. 1(b) and (a).}}
\end{figure}
\begin{figure*}[t]
{\includegraphics[width=1.8\columnwidth]{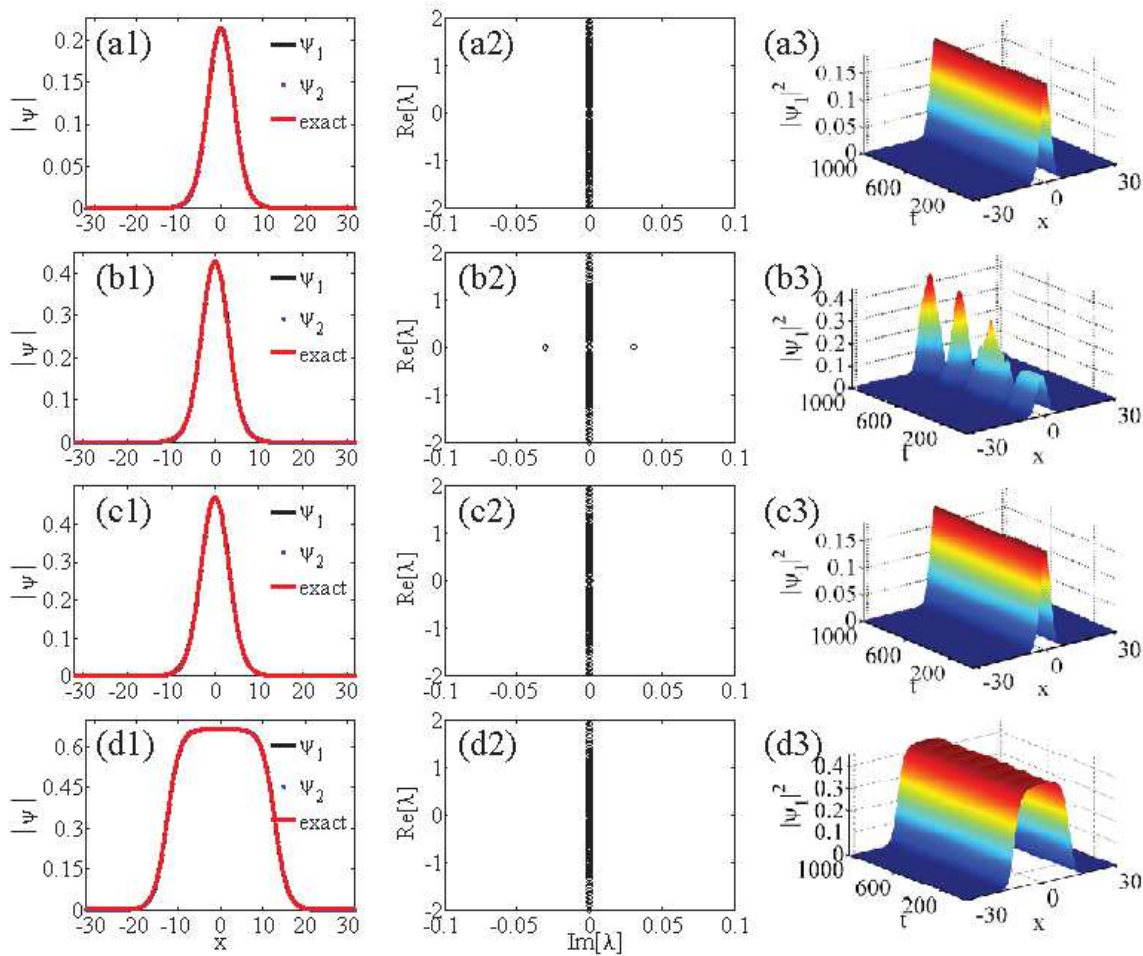}}
\caption{(Color online) The profiles of $|\psi_1(x)|$ and $|\psi_2(x)|$ components of the $\mathcal{PT}$-symmetric QDs, and the exact solution given by Eq. (\ref{exactsolution}), are shown by black solid, blue dotted, and red dashed curves, respectively, for different values of gain-loss parameter $\gamma$ and total norm $N$: (a1) $\gamma=0$ and $N=2$; (b1) $\gamma=0.03$ and $N=2$; (c1) $\gamma=0.03$ and $N=2.5$; (d1) $\gamma=0.03$ and $N=20$. The stability spectra of eigenvalues $\lambda$ for the corresponding $\mathcal{PT}$-symmetric QDs and the perturbed evolution of their $\psi_1$ component, are displayed in panels (a2)-(d2) and (a3)-(d3), respectively. The inter-core coupling strength is fixed as $\kappa=0.05$.}
\end{figure*}
\begin{figure*}[t]
{\includegraphics[width=1.8\columnwidth]{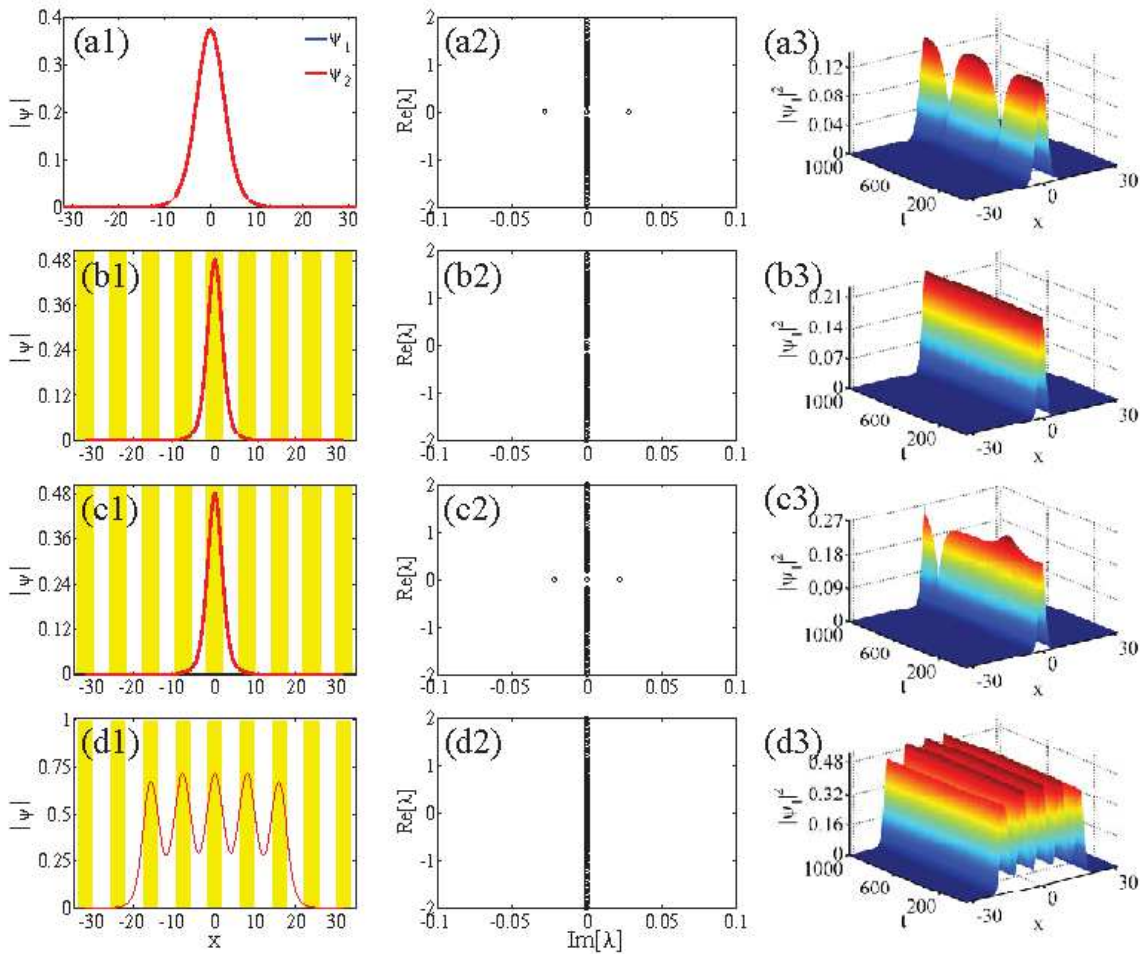}}
\caption{(Color online) Typical examples of the stability  of the $\mathcal{PT}$-symmetric on-site QDs in the presence of OL potential. (a1)-(a3) $V_0=0$, $\gamma=0$, and $N=1.5$; (b1)-(b3) $V_0=0.3$, $\gamma=0$, and $N=1.5$; (c1)-(c3) $V_0=0.3$, $\gamma=0.04$, and $N=1.5$; (d1)-(d3) $V_0=0.3$, $\gamma=0.04$, and $N=20$. The other parameters are set as $\kappa=0.05$, $D=8$, and $\theta=\pi/2$. The vertical yellow stripes denote the respective potential troughs in panels (b1)-(d1).}
\end{figure*}
\begin{figure*}[t]
{\includegraphics[width=1.8\columnwidth]{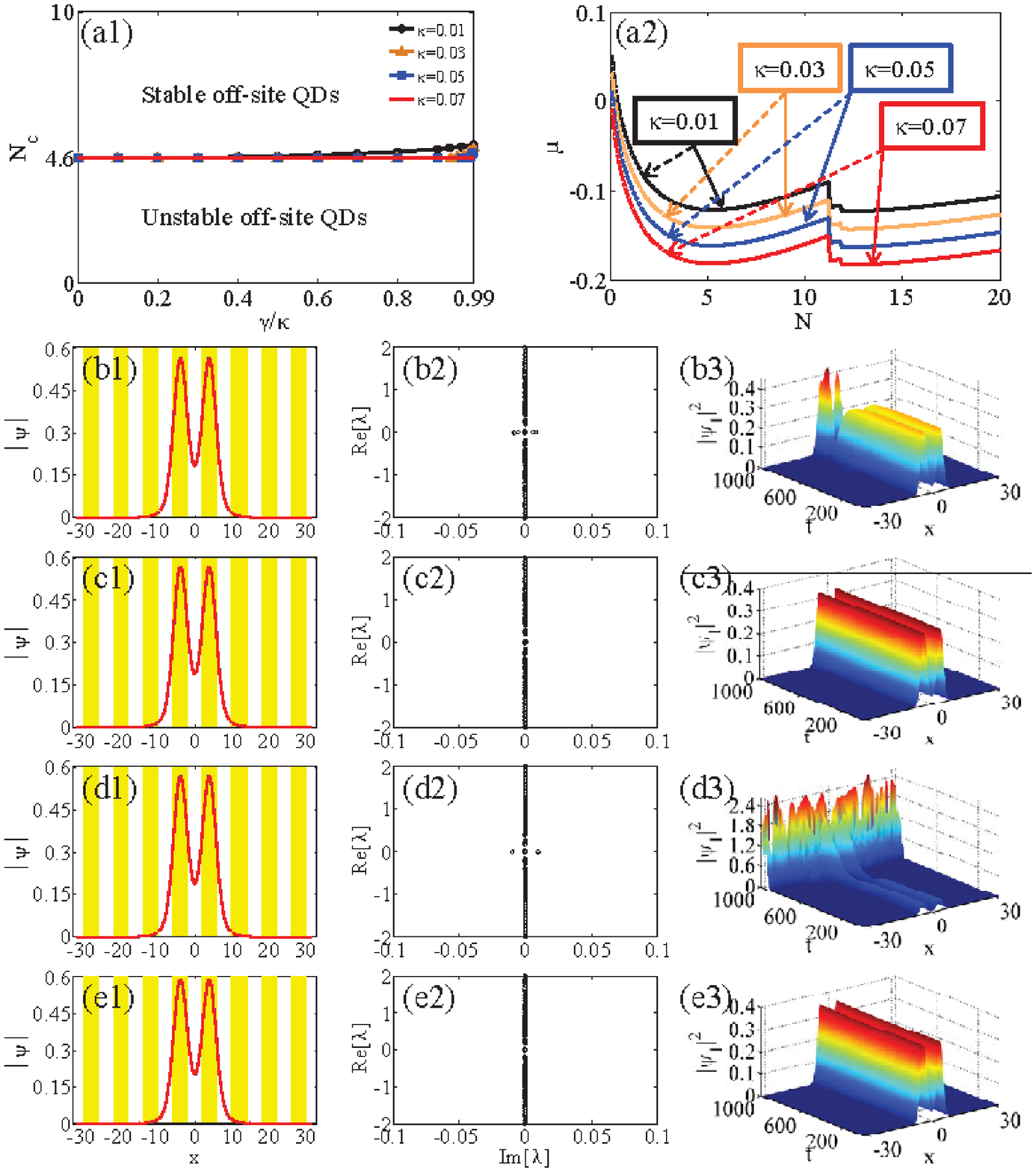}}
\caption{(Color online) (a1) Stability border for the $\mathcal{PT}$-symmetric off-site QDs with four different values of $\kappa$ ($\kappa$=0.01, 0.03, 0.05, and 0.07). The $\mathcal{PT}$-symmetric off-site QDs with total norm $N$ are stable at $N\geqslant N_c$, while they are unstable at $N<N_c$. (a2) The chemical potential $\mu$ versus $N$ for the symmetric off-site QDs with the above four different values of $\kappa$. The dashed and solid lines in the four $\mu(N)$ curves correspond to the unstable and stable symmetric off-site QDs, respectively. Typical examples of the stability of the $\mathcal{PT}$-symmetric off-site QDs in panels (b1)-(e3). (b1)-(b3) $\gamma/\kappa=0.2$ and $N=4.5$; (c1)-(c3) $\gamma/\kappa=0.2$ and $N=4.6$; (d1)-(d3) $\gamma/\kappa=0.9$ and $N=4.6$; (e1)-(e3) $\gamma/\kappa=0.9$ and $N=5$. The other parameters are set as $\kappa=0.01$, $D=8$, and $\theta=0$.}
\end{figure*}

\section{Stability analysis for $\mathcal{PT}$-symmetric quantum droplets}

\subsection{In the absence of optical lattice potential}
In reality the system cannot be perfectly isolated and it is expected that dynamics of the system are always robust against small fluctuations. The $\mathcal{PT}$-symmetric QDs may be observed experimentally only if perturbed QDs can survive for a sufficiently long time. Therefore, an important issue is to check the stability of the $\mathcal{PT}$-symmetric QDs.

We have performed numerical simulation if the evolution of the symmetric QDs against small symmetry breaking perturbations based on Eq. (\ref{GPEs}), aiming to identify stability boundaries for the symmetric QDs. Perturbations were introduced by adding 1\% random noise into the initial conditions. The stability of symmetric QDs in the $\mathcal{PT}$-symmetric system depends strongly on the values of gain-loss parameter $\gamma$, inter-core coupling strength $\kappa$, and total condensate norm $N$. The symmetric QDs are entirely unstable against the breaking of $\mathcal{PT}$ symmetry for $\gamma/\kappa\geqslant 1$. We plot the stability border for the symmetric QDs, showing $N$ as a function of $\gamma/\kappa$ for the different values of $\kappa$. The results are summarized in Fig. 2(a). For instance, for $\kappa=0.03$, it is found that the $\mathcal{PT}$-symmetric QDs are stable for the sufficiently small values of $N$. With the increase of $N$, the $\mathcal{PT}$-symmetric QDs become unstable, whereas they retrieve the stability at larger $N$. In other words, the $\mathcal{PT}$-symmetric QDs are unstable in the interval of $N_{c1}<N<N_{c2}$, and they are stable at $N<N_{c1}$ and $N>N_{c2}$. It is relevant to mention that, in the absence of the gain and loss ($\gamma=0$), the stable and unstable regions exactly correspond to those of the symmetric QDs in the conservative dual-core trap \cite{Liu053602}. The unstable region of the $\mathcal{PT}$-symmetric QDs becomes wider as the increase of $\gamma/\kappa$. As expected, the $\mathcal{PT}$ symmetry always break the stability of QDs. Branches $N_{c1}$ and $N_{c2}$ remain completely separated in the whole interval of $0\leqslant\gamma/\kappa\leqslant0.99$ for $\kappa<\kappa_c\approx 0.0592$. For $\kappa_c<\kappa\leqslant \kappa_{max}\approx 0.4$, the branches $N_{c1}$ and $N_{c2}$ partly merge and disappear at small values of $\gamma/\kappa$. This indicates that all the $\mathcal{PT}$-symmetric QDs, regardless of the value of $N$, are stable at $\gamma/\kappa<(\gamma/\kappa)_c$, with $(\gamma/\kappa)_c$ being the merger point of the $N_{c1}$ and $N_{c2}$ branches. By further increasing $\kappa$, all $\mathcal{PT}$-symmetric QDs, regardless of the value of $\gamma/\kappa$, are completely stable at $\kappa>\kappa_{max}\approx 0.4$. Accordingly, the $\mathcal{PT}$-symmetric QDs become unbreakable at $\kappa>\kappa_{max}$, viz., the blow up of $\mathcal{PT}$-symmetric QDs can be suppressed by the strong coupling strength between two cores.

The above predications for the stability and instability of $\mathcal{PT}$-symmetric QDs were verified by the linear stability analysis and direct simulations of the perturbed evolution of the $\mathcal{PT}$-symmetric QDs. The linear stability analysis for the stationary states was performed by adding small perturbations, $\psi_{1,2}=[\phi_{1,2}+u_{1,2}e^{i\lambda t}+v_{1,2}^\ast e^{-i\lambda^\ast t}]e^{-i\mu t}$, where $\phi_{1,2}$ are stationary wave functions with chemical potential $\mu$, $u_{1,2}$ and $v_{1,2}$ are perturbation eigenmodes, and $\lambda$ indicates the growth rate of the perturbation. Evidently, the $\mathcal{PT}$-symmetric QDs are unstable if $\lambda$ has an imaginary component, while they are stable if $\lambda$ is real. Fig. 3 displays typical examples of the stable and unstable symmetric QDs with different values of the total norm $N$ for a fixed hopping rate, e.g., $\kappa=0.05$. Fig. 3(a1) shows the wave-function profiles of Guassian-shaped symmetric QDs corresponding to a moderate value of the norm ($N=2$) without gain-loss parameter ($\gamma=0$). The symmetric QDs, which are expected to be stable based on the above analysis in Fig. 1(a) ($\gamma=0$), indeed remains stable [see Figs. 3(a2) and 3(a3)]. The result is consistent with that of QDs in the conservative dual-core trap \cite{Liu053602}. The presence of $\mathcal{PT}$-symmetric potential modifies this physical picture. To show the effect of the $\mathcal{PT}$ symmetry on QDs, we choose a fixed gain-loss parameter $\gamma=0.03$, for different values of $N$, such as $N=2$, $2.5$, and $20$, as can be seen in Figs. 3(b1)-(d1), respectively. Figs. 3(b1)-(b3) clearly shows $\mathcal{PT}$ symmetry breaks the stability of QDs. However, the stability of $\mathcal{PT}$-symmetric QDs can be restored for relatively large values of $N$ provided that $N$ belongs to either of the above predicted stability areas ($N>N_{c2}$), as shown in Figs. 3(c1)-(c3). Particularly, the $\mathcal{PT}$-symmetric QDs can also display a typical broad flat-top profile for the large droplets $N=20$ in Figs. 3(d1)-(d3).

\subsection{In the presence of optical lattice potential}
Next, we will deal with the effect of the interplay between the OL potential and $\mathcal{PT}$ symmetry on the stability of the symmetric QDs. The OL potential $V_0\cos^2(\frac{\pi}{D}x+\theta)$ is added into Eq. (\ref{GPEs}), where the parameters $V_0$, $D$, and $\theta$ denote depth, period, and phase of OL potential, respectively. Without loss of generality, the period of OL potential is fixed as $D=8$ throughout the present work. We start by considering the stability of on-site QDs ($\theta=0$) in the $\mathcal{PT}$-symmetric dual-core couplers. Following the procedure of the above numerical analysis, we identify stability areas for symmetric on-site QDs, as can be seen in Fig. 2(b). Similar to the situation in the absence of OL potential, there are still two stability areas, $N<N_{c1}$ and $N>N_{c2}$, for the symmetric on-site QDs. However, the two threshold values, viz., $N_{c1}$ and $N_{c2}$, become smaller by comparing Figs. 2(b) and (a). This indicates that the OL potential can assist stabilization of symmetric on-site QDs for moderate values of $N$. Fixing $\kappa=0.05$, we display typical examples of the stable and unstable $\mathcal{PT}$-symmetric on-site QDs in Fig. 4. For a characteristic moderate value of norm $N=1.5$, Figs. 4(a1-a3) reveal that such symmetric QDs are unstable at $V_0=0$ and $\gamma=0$ [also see Fig. 2(a)], whereas it becomes stable with the assistance of the OL potential. With the increase of gain and loss $\gamma$, $\mathcal{PT}$-symmetric potential will always break the stability of the on-site QDs [e.g., see Figs. 4(c1)-(c3)]. As expected, Figs. 4(d1)-(d3) confirm that the $\mathcal{PT}$-symmetric on-site QDs are stable for large $N$ in the above predicted stability areas.

The situation is obviously different for $\mathcal{PT}$-symmetric off-site QDs ($\theta=0$). We scanned a broad range of $N$ to analyze the stability of off-site QDs in the $\mathcal{PT}$-symmetric potential for different values of $\kappa$. The numerical results indicate that there exists a stability border in the total condensate norm $N$. The $\mathcal{PT}$-symmetric off-site QDs are unstable at $N<N_c$, while they become stable at $N\geqslant N_c$. Fig. 5(a1) shows the stability border $N_c$ as a function of $\gamma/\kappa$ for different values of $\kappa$. It is seen that, in the absence of the gain and loss ($\gamma=0$), the critical point for the unstable and stable symmetric off-site QDs is $N_c(\gamma=0)\approx 4.6$. To explore physical mechanism for the formation of stable symmetric off-site QDs, in Fig. 5(a2) we plot the chemical potential $\mu$ as a function of $N$ for the four different values of $\kappa$. The dashed and solid lines in the four $\mu(N)$ curves correspond to the unstable and stable symmetric off-site QDs, respectively. Note that the chemical potential $\mu$ approximately corresponds to the respective minimal value at the critical point $N=4.6$. Due to the dominating cubic self-repulsion with the competition of the quadratic self-attraction in the relatively large off-site QDs, the $\mu(N)$ curves satisfy the necessary stability condition in the form of the anti-Vakhiton-Kolokolov criterion in repulsive nonlinear interaction \cite{Sakaguchi013624}, $d\mu/dN>0$, except for suddenly changing at some special values of $N$. The results are similar to that of the single-component off-site QDs in OL with $N=2.3$ for this case \cite{Zhou104881}. For the weakly-couple case ($\kappa<\kappa_{max}\approx 0.07$), the $\mathcal{PT}$ symmetry tends to break the stability of off-site QDs with the increase of $\gamma/\kappa$, as shown in Fig. 5(a1). The $\mathcal{PT}$-symmetric off-site QDs retrieve the stability at still larger values of $N$. Above a certain value of the hopping strength ($\kappa\geqslant\kappa_{max}\approx 0.07$), the $\mathcal{PT}$-symmetric off-site QDs, which are no longer dependent of the gain-loss parameter $\gamma$, are unstable for $N<4.6$, while they become entire stable for $N\geqslant 4.6$. Accordingly, by fixing $V_0=0.3$ and $\kappa=0.01$, typical examples of stable and unstable $\mathcal{PT}$-symmetric off-site QDs located close to the stability border for different values of $N$, are displayed in Figs. 5(b1)-(e1). For $\gamma/\kappa=0.2$, the $\mathcal{PT}$-symmetric off-site QDs corresponding to $N=4.5<N_c$ is unstable, while increasing to $N=4.6$ it becomes stable, as shown in Figs. 5(b1)-(b3) and Figs. 5(c1)-(c3), respectively. However, for a bigger $\gamma/\kappa$, such as $\gamma/\kappa=0.9$, the $\mathcal{PT}$-symmetric off-site QDs corresponding to $N=4.6$ is no longer stable, as shown in Figs. 5(d1)-(d3). Figs. 5(e1)-(e3) indicates that the larger $\mathcal{PT}$-symmetric off-site QDs (e.g., $N=5$) can retrieve the stability. As expected, these results are consistent with those in Fig. 5(a1).

\section{Collisions of $\mathcal{PT}$-symmetric quantum droplets}
Because the presence of the gain and loss does not break the Galilean invariance of the underlying equation (\ref{GPEs}), it is relevant to explore collisions between the moving $\mathcal{PT}$-symmetric QDs. In the conservative dual-core trap, it has been demonstrated that the colliding QDs tend to merger into breathers unless they move very fast \cite{Liu053602}.

We simulated the collisions, and the corresponding initial conditions for Eq. (\ref{GPEs}) were taken as
\begin{equation}\label{collisions}
\psi_{1,2}(x,t=0)=\psi(x+x_0)e^{ikx}+\psi(x-x_0)e^{-ikx},
\end{equation}
where $\psi(x)$ represents the stationary shape of symmetric QDs, and $k$ is a kick determining the velocity of QDs. This ansatz approximates a solution comprising two initial QDs located at $-x_0$ and $x_0$. In the simulations we varied $k$ and $N$.

\begin{figure}[htp] \center
\includegraphics[width=3.6in]{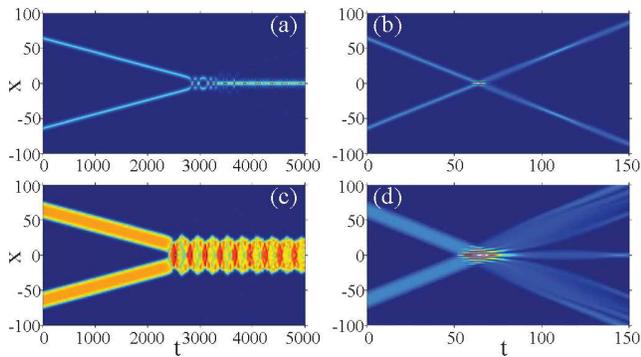}
\caption{\scriptsize{(Color online) Typical examples of density plots, in component $\psi_1$, for collisions between two $\mathcal{PT}$-symmetric QDs in the absence of OL potential, launched as per Eq. (7) with $x_0=64$. (a) $N=1$ and $k=0.02$; (b) $N=1$ and $k=1$; (c) $N=20$ and $k=0.02$; (d) $N=20$ and $k=1$. The other parameters are fixed as $\kappa=0.08$ and $\gamma/\kappa=0.3$.}}
\end{figure}

In the absence of OL potential ($V_0=0$), Fig. 6 shows typical collision pictures for small and large $\mathcal{PT}$-symmetric QDs located in the stable region, corresponding to $N=1$ and $N=20$, respectively, for $\kappa=0.08$ and $\gamma/\kappa=0.3$. It is observed that the slowly moving $\mathcal{PT}$-symmetric QDs merge into breathers after the collision at relatively small values of $k$, as shown in Figs. 6(a) and (c). With the increase of $k$, fast-moving small $\mathcal{PT}$-symmetric QDs pass through each other, i.e., the quasi-elastic collision occurs in this case (see Fig. 6(b)). The situation is quite different for large $\mathcal{PT}$-symmetric QDs. In this case, the shapes of the QDs are no longer preserved after the collision, and they undergo fragmentation resulting in the formation of three outgoing QDs with a majority of particles being kept in the moving ones and forming a small quiescent one (see Fig. 6(d)). The similar results can also be observed in the presence of OL potential, as shown in Fig. 7. The slowly moving $\mathcal{PT}$-symmetric QDs in shallow OL potential tend to merge into breathers after the collision [see Figs. 7(a) and (c)]. Fast-moving small $\mathcal{PT}$-symmetric QDs pass through each other quasi-elastically, while fast-moving large $\mathcal{PT}$-symmetric QDs collide inelastically and undergo fragmentation in OL potential, as shown in Figs. 7(b) and (c), respectively.

\begin{figure}[htp] \center
\includegraphics[width=3.6in]{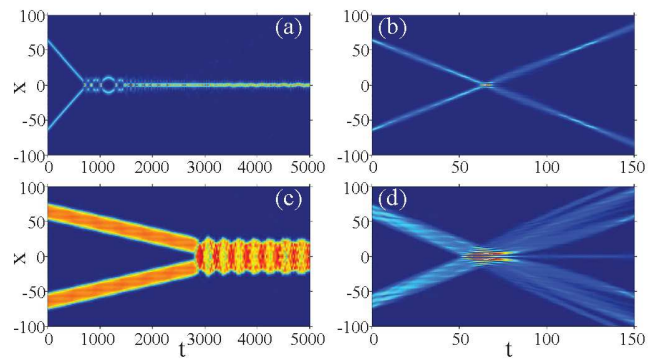}
\caption{\scriptsize{(Color online) Typical examples of density plots for collisions between two $\mathcal{PT}$-symmetric on-site QDs in the presence of OL potential. (a) $N=1$, $k=0.1$, and $V_0=0.01$; (b) $N=1$, $k=1$, and $V_0=0.1$; (c) $N=20$, $k=0.02$, and $V_0=0.01$; (d) $N=20$, $k=1$, and $V_0=0.1$. The other parameters are fixed as $x_0=64$, $\kappa=0.08$, $\gamma/\kappa=0.3$, $D=8$, and $\theta=\pi/2$.}}
\end{figure}

\section{conclusion and discussion}
The objective of this work is to extend the study of the spontaneous symmetry breaking of quantum droplets (QDs) in the dual-core trap \cite{Liu053602}, to dynamics of QDs in parity-time ($\mathcal{PT}$)-symmetric dual-core setting. Such $\mathcal{PT}$-symmetric system cannot support stable asymmetric QDs as the balance between the gain and loss is impossible for them. Therefore, the main subject is the effect of the interplay between $\mathcal{PT}$-symmetric potential and optical lattice (OL) potential on the stability and collisions of symmetric QDs in the unbroken $\mathcal{PT}$-symmetric phase. It is found that $\mathcal{PT}$-symmetric QDs in the absence of OL potential and $\mathcal{PT}$-symmetric on-site QDs display similar stability condition, i.e., both of them are stable for relatively small and large values of the total condensate norm $N$, and are unstable for moderate values of $N$. The difference is the OL potential can assist stabilization of $\mathcal{PT}$-symmetric on-site QDs for some moderate values of $N$. However, the situation is obviously different for $\mathcal{PT}$-symmetric off-site QDs, and only the relatively large $\mathcal{PT}$-symmetric off-site QDs are stable. Finally, collisions between stable $\mathcal{PT}$-symmetric QDs are systematically studied both in the absence of OL potential and in the presence of OL potential too. It is revealed that the collision dynamics of the stable $\mathcal{PT}$-symmetric QDs depend on the velocity and the total condensate norm $N$ of QDs. The slowly moving $\mathcal{PT}$-symmetric QDs regardless of the value of $N$ tend to merge into breathers. The small fast-moving $\mathcal{PT}$-symmetric QDs is quasi-elastic, while the large fast-moving ones suffer fragmentation after the collision.

It is worth noting that in the present work we extend and generalize the previous study of dynamics of QDs in both dual-core trap \cite{Liu053602} and optical lattice \cite{Zhou104881} to $\mathcal{PT}$-symmetric dual-core setting, and there exist some similarities between them. However, the generalization is nontrivial. On the one hand, the finding of $\mathcal{PT}$ symmetry opened the possibility of postulating new theoretical concepts to replace the long accepted requirement of Hermitian Hamiltonians \cite{Bender270401}, and the studies of $\mathcal{PT}$ symmetry-based effects revealed many interesting phenomena, including nonreciprocal light propagation \cite{Ruter192}, unidirectional invisibility \cite{Lin213901}, topological bound state \cite{Weimann433}, and anomalous edge states \cite{Rivolta023864}. On the other hand, generalizing the notion of $\mathcal{PT}$ symmetry to the QD case is still lacking to date. Therefore, combining $\mathcal{PT}$ symmetry with QDs, we can broaden the physical contents of QDs and lay the foundation for further combination and research.

\begin{acknowledgments}
This work is supported by the National Natural Science Foundation of China under Grant Nos. 11805283. The Hunan Provincial Natural Science Foundation under Grant Nos. 2019JJ40060 and 2019JJ30044. The innovation platform open fund project of Hengyang Normal University under Grant No. GD20K02. The aid program of science and technology innovative research team of Hunan Institute of Technology under Grant No. TD18001. The construct program of the key discipline in Hunan Institute of Technology under Grant No. XK19010.
\end{acknowledgments}


\begin{thebibliography}{999}

\bibitem{Baillie255302} D. Baillie, R. M. Wilson, and P. B. Blakie, Collective Excitations of Self-Bound Droplets of a Dipolar Quantum Fluid, Phys. Rev. Lett. {\bf 119}, 255302 (2017).
\bibitem{Wachtler061603} F. W\"{a}chtler and L. Santos, Quantum filaments in dipolar Bose-Einstein condensates, Phys. Rev. A {\bf 93}, 061603(R) (2016).
\bibitem{Baillie021602} D. Baillie, R. M. Wilson, R. N. Bisset, and P. B. Blakie, Self-bound dipolar droplet: A localized matter wave in free space, Phys. Rev. A {\bf 94}, 021602(R) (2016).
\bibitem{Barbut160402} I. Ferrier-Barbut, M. Wenzel, F. B\"{o}ttcher, T. Langen, M. Isoard, S. Stringari, and T. Pfau, Scissors Mode of Dipolar Quantum Droplets of Dysprosium Atoms, Phys. Rev. Lett. {\bf 120}, 160402 (2018).
\bibitem{Ferioli090401} G. Ferioli, G. Semeghini, L. Masi, G. Giusti, G. Modugno, M. Inguscio, A. Gallem\'{i}, A. Recati, and M. Fattori, Collisions of Self-Bound Quantum Droplets, Phys. Rev. Lett. {\bf 122}, 090401 (2019).
\bibitem{Kartashov193902} Y. V. Kartashov, B. A. Malomed, and L. Torner, Metastability of Quantum Droplet Clusters, Phys. Rev. Lett. {\bf 122}, 193902 (2019).
\bibitem{Tengstrand160405} M. N. Tengstrand, P. St\"{u}rmer, E. \"{O}. Karabulut, and S. M. Reimann, Rotating Binary Bose-Einstein Condensates and Vortex Clusters in Quantum Droplets, Phys. Rev. Lett. {\bf 123}, 160405 (2019).
\bibitem{Errico033155} C. D'Errico, A. Burchianti, M. Prevedelli, L. Salasnich, F. Ancilotto, M. Modugno, F. Minardi, and C. Fort, Observation of Quantum Droplets in a Heteronuclear Bosonic Mixture, Phys. Rev. Research {\bf 1}, 033155 (2019).
\bibitem{Chiquillo051601} E. Chiquillo, Low-dimensional self-bound quantum Rabi-coupled bosonic droplets, Phys. Rev. A {\bf 99}, 051601(R) (2019).
\bibitem{Mishra073402} C. Mishra, L. Santos, and R. Nath, Self-Bound Doubly Dipolar Bose-Einstein Condensates, Phys. Rev. Lett. {\bf 124}, 073402 (2020).
\bibitem{Ferioli013269} G. Ferioli, G. Semeghini, S. Terradas-Brians\'{o}, L. Masi, M. Fattori, and M. Modugno, Dynamical formation of quantum droplets in a $^{39}$K mixture, Phys. Rev. Research {\bf 2}, 013269 (2020).
\bibitem{Oldziejewski090401} R. O{\l}dziejewski, W. G\'{o}recki, K. Paw{\l}owski, and K. Rza\.{z}ewski, Strongly Correlated Quantum Droplets in Quasi-1D Dipolar Bose Gas, Phys. Rev. Lett. {\bf 124}, 090401 (2020).
\bibitem{Zhang133901} X. Zhang, X. Xu, Y. Zheng, Z. Chen, B. Liu, C. Huang, B. A. Malomed, and Y. Li, Semidiscrete Quantum Droplets and Vortices, Phys. Rev. Lett. {\bf 123}, 133901 (2020).

\bibitem{Petrov155302} D. S. Petrov, Quantum Mechanical Stabilization of a Collapsing Bose-Bose Mixture, Phys. Rev. Lett. {\bf 115}, 155302 (2015).
\bibitem{Lee1135} T. D. Lee, K. Huang, and C. N. Yang, Eigenvalues and Eigenfunctions of a Bose System of Hard Spheres and Its Low-Temperature Properties, Phys. Rev. {\bf 106}, 1135 (1957).

\bibitem{Griesmaier160401} A. Griesmaier, J. Werner, S. Hensler, J. Stuhler, and T. Pfau, Bose-Einstein Condensation of Chromium, Phys. Rev. Lett. {\bf 94}, 160401 (2005).
\bibitem{Beaufils061601} Q. Beaufils, R. Chicireanu, T. Zanon, B. Laburthe-Tolra, E. Mar\'{e}chal, L. Vernac, J. -C. Keller, and O. Gorceix, All-optical production of chromium Bose-Einstein condensates, Phys. Rev. A {\bf 77}, 061601(R) (2008).

\bibitem{Kadau194} H. Kadau, M. Schmitt, M. Wentzel, C. Wink, T. Maier, I. Ferrier-Barbut, and T. Pfau, Observing the Rosenzweig instability of a quantum ferrofluid, Nature {\bf 530}, 194 (2016).
\bibitem{Barbut215301} I. Ferrier-Barbut, H. Kadau, M. Schmitt, M. Wenzel, and T. Pfau, Observation of Quantum Droplets in a Strongly Dipolar Bose Gas, Phys. Rev. Lett. {\bf 116}, 215301 (2016).
\bibitem{Schmitt259} M. Schmitt, M. Wenzel, F. B\"{o}ttcher, I. Ferrier-Barbut, and T. Pfau, Self-bound droplets of a dilute magnetic quantum liquid, Nature {\bf 539}, 259 (2016).
\bibitem{Chomaz041039} L. Chomaz, S. Baier, D. Petter, M. J. Mark, F. W\"{a}chtler, L. Santos, and F. Ferlaino, Quantum-Fluctuation-Driven Crossover from a Dilute Bose-Einstein Condensate to a Macrodroplet in a Dipolar Quantum Fluid, Phys. Rev. X {\bf 6}, 041039 (2016).
\bibitem{Saito053001} H. Saito, Path-integral Monte-Carlo study on a droplet of a dipolar Bose-Einstein condensate stabilized by quantum fluctuation, J. Phys. Soc. Jpn. {\bf 85}, 053001 (2016).

\bibitem{Cheiney135301} P. Cheiney, C. R. Cabrera, J. Sanz, B. Naylor, L. Tanzi, and L. Tarruell, Bright Soliton to Quantum Droplet Transition in a Mixture of Bose-Einstein Condensates, Phys. Rev. Lett. {\bf 120}, 135301 (2018).
\bibitem{Cabrera301} C. R. Cabrera, L. Tanzi, J. Sanz, B. Naylor, P. Thomas, P. Cheiney, and L. Tarruell, Quantum liquid droplets in a mixture of Bose-Einstein condensates, Science {\bf 359}, 301 (2018).
\bibitem{Semeghini235301} G. Semeghini, G. Ferioli, L. Masi, C. Mazzinghi, L. Wolswijk, F. Minardi, M. Modugno, G. Modugno, M. Inguscio, and M. Fattori, Self-Bound Quantum Droplets of Atomic Mixtures in Free Space, Phys. Rev. Lett. {\bf 120}, 235301 (2018).

\bibitem{Petrov100401} D. S. Petrov and G. Astrakharchik, Ultradilute Low-Dimensional Liquids, Phys. Rev. Lett. {\bf 117}, 100401 (2016).

\bibitem{Li113043} Y. Li, Z. Luo, Y. Liu, Z. Chen, C. Huang, S. Fu, H. Tan, and B. A. Malomed, Two-dimensional solitons and quantum droplets supported by competing self- and cross-interactions in spin-orbit-coupled condensates, New J. Phys. {\bf 19}, 113043 (2017); Y. Li, Z. Chen, Z. Luo, C. Huang, H. Tan, W. Pang, and B. A. Malomed, Two-dimensional vortex quantum droplets, Phys. Rev. A {\bf 98}, 063602 (2018).

\bibitem{Astrakharchik013631} G. E. Astrakharchik and B. A. Malomed, Dynamics of one-dimensional quantum droplets, Phys. Rev. A {\bf 98}, 013631 (2018).
\bibitem{Zhou104881} Z. Zhou, X. Yu, Y. Zou, and H. Zhong, Dynamics of quantum droplets in a one-dimensional optical lattice, Commun. Nonlinear. Sci. Numer. Simulat. {\bf 78}, 104881 (2019).
\bibitem{Liu053602} B. Liu, H. Zhang, R. Zhong, X. Zhang, X. Qin, C. Huang, Y. Li, and B. A. Malomed, Symmetry breaking of quantum droplets in a dual-core trap, Phys. Rev. A {\bf 99}, 053602 (2019).

\bibitem{Bender5243} C. M. Bender and S. Boettcher, Real Spectra in Non-Hermitian Hamiltonians Having $\mathcal{PT}$ Symmetry, Phys. Rev. Lett. {\bf 80}, 5243 (1998).
\bibitem{Bender947} C. M. Bender, Making sense of non-Hermitian Hamiltonian, Rep. Prog. Phys. {\bf 70}, 947 (2007).

\bibitem{Longhi243} S. Longhi, Quantum-optical analogies using photonic structures, Laser and Photon. Rev. {\bf 3}, 243 (2008).
\bibitem{Makris103904} K. G. Makris, R. El-Ganainy, D. N. Christodoulides, and Z. H. Musslimani, Beam Dynamics in $\mathcal{PT}$ Symmetric Optical Lattices, Phys. Rev. Lett. {\bf 100}, 103904 (2008).
\bibitem{Musslimani030402} Z. H. Musslimani, K. G. Makris, R. El-Ganainy, and D. N. Christodoulides, Optical Solitons in $\mathcal{PT}$ Periodic Potentials, Phys. Rev. Lett. {\bf 100}, 030402 (2008).

\bibitem{Guo093902} A. Guo, G. J. Salamo, D. Duchesne, R.Morandotti, M. Volatier-Ravat, V. Aimez, G. A. Siviloglou, and D. N. Christodoulides, Observation of $\mathcal{PT}$-Symmetry Breaking in Complex Optical Potentials, Phys. Rev. Lett. {\bf 103}, 093902 (2009).
\bibitem{Ruter192} C. E. R\"{u}ter, K. G. Makris, R. El-Ganainy, D. N. Christodoulides, M. Segev, and D. Kip, Observation of parity-time symmetry in optics, Nat. Phys. {\bf 6}, 192 (2010).

\bibitem{Klaiman080402} S. Klaiman, U. G\"{u}nther, and N. Moiseyev, Visualization of Branch Points in $\mathcal{PT}$-Symmetric Waveguides, Phys. Rev. Lett. {\bf 101}, 080402 (2008).
\bibitem{Single042123} F. Single, H. Cartarius, G. Wunner, and J. Main, Coupling approach for the realization of a $\mathcal{PT}$-symmetric potential for a Bose-Einstein condensate in a double well, Phys. Rev. A {\bf 90}, 042123 (2014).
\bibitem{Gutohrlein335302} R. Gut\"{o}hrlein, J. Schnabel, I. Iskandarov, H. Cartarius, J. Main, and G. Wunner, Realizing $\mathcal{PT}$-symmetric BEC subsystems in closed Hermitian systems, J. Phys. A {\bf 48}, 335302 (2015).
\bibitem{Kreibich051601} M. Kreibich, J. Main, H. Cartarius, and G. Wunner, Hermitian four-well potential as a realization of a $\mathcal{PT}$-symmetric system, Phys. Rev. A {\bf 87}, 051601(R) (2013).
\bibitem{Kreibich033630} M. Kreibich, J. Main, H. Cartarius, and G. Wunner, Realizing $\mathcal{PT}$-symmetric non-Hermiticity with ultracold atoms and Hermitian multiwell potentials, Phys. Rev. A {\bf 90}, 033630 (2014).
\bibitem{Kreibich023624} M. Kreibich, J. Main, H. Cartarius, and G. Wunner, Tilted optical lattices with defects as realizations of $\mathcal{PT}$ symmetry in Bose-Einstein condensates, Phys. Rev. A {\bf 93}, 023624 (2016).
\bibitem{Kogel063610} F. Kogel, S. Kotzur, D. Dizdarevic, J. Main, and G. Wunner, Realization of $\mathcal{PT}$-symmetric and $\mathcal{PT}$-symmetry-broken states in static optical-lattice potentials, Phys. Rev. A {\bf 99}, 063610 (2019).

\bibitem{Driben4323} R. Driben and B. A. Malomed, Stability of solitons in parity-time-symmetric couplers, Opt. Lett. {\bf 36}, 4323 (2011).
\bibitem{Peng394} B. Peng, S. K. Ozdemir, F. Lei, F. Monifi, M. Gianfreda, G. Long, S. Fan, F. Nori, C. M. Bender, and L. Yang, Parity-time-symmetric whispering-gallery microcavities, Nat. Phys. {\bf 10}, 394 (2014).

\bibitem{Bittner024101} S. Bittner, B. Dietz, U. G\"{u}nther, H. L. Harney, M. Miski-Oglu, A. Richter, and F. Sch\"{a}fer, $\mathcal{PT}$ Symmetry and Spontaneous Symmetry Breaking in a Microwave Billiard, Phys. Rev. Lett. {\bf 108}, 024101 (2012).
\bibitem{Huai043803} S. Huai, Y. Liu, J. Zhang, L. Yang, and Y. Liu, Enhanced sideband responses in a $\mathcal{PT}$-symmetric-like cavity magnomechanical system, Phys. Rev. A {\bf 99}, 043803 (2019).

\bibitem{Bender234101} N. Bender, S. Factor, J. D. Bodyfelt, H. Ramezani, D. N. Christodoulides, F. M. Ellis, and T. Kottos, Observation of Asymmetric Transport in Structures with Active Nonlinearities, Phys. Rev. Lett. {\bf 110}, 234101 (2013).
\bibitem{Bender040101} N. Bender, S. Factor, J. D. Bodyfelt, H. Ramezani, D. N. Christodoulides, F. M. Ellis, and T. Kottos, Experimental study of active LRC circuits with $\mathcal{PT}$ symmetries, Phys. Rev. A {\bf 84}, 040101(R) (2011).

\bibitem{Li855} J. Li, A. K. Harter, J. Liu, L. de Melo, Y. N. Joglekar, and L. Luo, Observation of parity-time symmetry breaking transitions in a dissipative Floquet system of ultracold atoms, Nat. Commun. {\bf 10}, 855 (2019).

\bibitem{Cartarius013612} H. Cartarius and G. Wunner, Model of a $\mathcal{PT}$-symmetric Bose-Einstein condensate in a $\delta$-function double-well potential, Phys. Rev. A {\bf 86}, 013612 (2012).
\bibitem{Dast124} D. Dast, D. Haag, H. Cartarius, G. Wunner, R. Eichler, and J. Main, A Bose-Einstein condensate in a $\mathcal{PT}$ symmetric double well, Fortschr. Phys. {\bf 61}, 124 (2013).


\bibitem{Haag023601} D. Haag, D. Dast, A. L\"{o}hle, H. Cartarius, J. Main, and G. Wunner, Nonlinear quantum dynamics in a $\mathcal{PT}$-symmetric double well, Phys. Rev. A {\bf 89}, 023601 (2014).
\bibitem{Fortanier063608} R. Fortanier, D. Dast, D. Haag, H. Cartarius, J. Main, G. Wunner, and R. Gut\"{o}hrlein, Dipolar Bose-Einstein condensates in a $\mathcal{PT}$-symmetric double-well potential, Phys. Rev. A {\bf 89}, 063608 (2014).
\bibitem{Dast033617} D. Dast, D. Haag, H. Cartarius, and G. Wunner, Purity oscillations in Bose-Einstein condensates with balanced gain and loss, Phys. Rev. A {\bf 93}, 033617 (2016).

\bibitem{Dast053601} D. Dast, D. Haag, H. Cartarius, J. Main, and G. Wunner, Bose-Einstein condensates with balanced gain and loss beyond mean-field theory, Phys. Rev. A {\bf 94}, 053601 (2016).
\bibitem{Dast023625} D. Dast, D. Haag, H. Cartarius, J. Main, and G. Wunner, Stationary states in the many-particle description of Bose-Einstein condensates with balanced gain and loss, Phys. Rev. A {\bf 96}, 023625 (2017).
\bibitem{Lunt023614} P. Lunt, D. Haag, D. Dast, H. Cartarius, and G. Wunner, Balanced gain and loss in Bose-Einstein condensates without $\mathcal{PT}$ symmetry, Phys. Rev. A {\bf 96}, 023614 (2017).
\bibitem{Haag033607} D. Haag, D. Dast, H. Cartarius, and G. Wunner, $\mathcal{PT}$-symmetric gain and loss in a rotating Bose-Einstein condensate, Phys. Rev. A {\bf 97}, 033607 (2018).
\bibitem{Zhou043412} Z. Zhou and Z. Yu, Interaction effects on the $\mathcal{PT}$-symmetry-breaking transition in atomic gases, Phys. Rev. A {\bf 99}, 043412 (2019).

\bibitem{Li013604} Y. Li, J. Liu, W. Pang, and B. A. Malomed, Symmetry breaking in dipolar matter-wave solitons in dual-core couplers, Phys. Rev. A {\bf 87}, 013604 (2013).

\bibitem{Burlak113103} G. Burlak, S. Garcia-Paredes, and B. A. Malomed, $\mathcal{PT}$-symmetric couplers with competing cubic-quintic nonlinearities, Chaos {\bf 26}, 113103 (2016).
\bibitem{Burlak062904} G. Burlak and B. A. Malomed, Stability boundary and collisions of two-dimensional solitons in $\mathcal{PT}$-symmetric couplers with the cubic-quintic nonlinearity, Phys. Rev. A {\bf 88}, 062904 (2013).

\bibitem{Zakrzewski3748} J. Zakrzewski, Analytic solutions of the two-state problem for a class of chirped pulses, Phys. Rev. A {\bf 32}, 3748 (1985).
\bibitem{Dunlap3625} D. H. Dunlap and V. M. Kenkre, Dynamic localization of a charged particle moving under the influence of an electric field, Phys. Rev. B {\bf 34}, 3625 (1986).

\bibitem{Kato1966} T. Kato, Perturbation Theroy of Linear Operators (Springer-Verlag, Berlin, New York, 1966).
\bibitem{Heiss2455} W. D. Heiss, Exceptional points of non-Hermitian operators, J. Phys. A {\bf 37}, 2455 (2004).

\bibitem{Cao61} H. Cao and J. Wiersig, Dielectric microcavities: Model systems for wave chaos and non-Hermitian physics, Rev. Mod. Phys. {\bf 87}, 61 (2015).
\bibitem{Heiss444016} W. D. Heiss, The physics of exceptional points, J. Phys. A {\bf 45}, 444016 (2012).
\bibitem{Lee133903} T. E. Lee, Anomalous Edge State in a Non-Hermitian Lattice, Phys. Rev. Lett. {\bf 116}, 133903 (2016).
\bibitem{Leykam040401} D. Leykam, K. Y. Bliokh, C. Huang, Y. D. Chong, and F. Nori, Edge Modes, Degeneracies, and Topological Numbers in Non-Hermitian Systems, Phys. Rev. Lett. {\bf 118}, 040401 (2017).
\bibitem{Shen146402} H. Shen, B. Zhen, and L. Fu, Topological Band Theory for Non-Hermitian Hamiltonians, Phys. Rev. Lett. {\bf 120}, 146402 (2018).

\bibitem{Hassan093002} A. U. Hassan, B. Zhen, M. Solja\v{c}i\'{c}, M. Khajavikhan, and D. N. Christodoulides, Dynamically Encircling Exceptional Points: Exact Evolution and Polarization State Conversion, Phys. Rev. Lett. {\bf 118}, 093002 (2017).
\bibitem{Doppler76} J. Doppler, A. A. Mailybaev, J. B\"{o}hm, U. Kuhl, A. Girschik, F. Libisch, T. J. Milburn, P. Rabl, N. Moiseyev, and S. Rotter, Dynamically encircling an exceptional point for asymmetric mode switching, Nature (London) {\bf 537}, 76 (2016).
\bibitem{Xu80} H. Xu, D. Mason, L. Jiang, and J. G. E. Harris, Topological energy transfer in an optomechanical system with exceptional points, Nature (London) {\bf 537}, 80 (2016).


\bibitem{Makris063807} K. G. Makris, R. El-Ganainy, D. N. Christodoulides, and Z. H. Musslimani, $\mathcal{PT}$-symmetric optical lattices, Phys. Rev. A {\bf 81}, 063807 (2010).
\bibitem{Nixon023822} S. Nixon, L. Ge, and J. Yang, Stability analysis for solitons in $\mathcal{PT}$-symmetric optical lattices, Phys. Rev. A {\bf 85}, 023822 (2012).

\bibitem{Sakaguchi013624} H. Sakaguchi and B. A. Malomed, Solitons in combined linear and nonlinear lattice potentials, Phys. Rev. A {\bf 81}, 013624 (2010).

\bibitem{Bender270401} C. M. Bender, D. C. Brody, and H. F. Jones, Complex Extension of Quantum Mechanics, Phys. Rev. Lett. {\bf 89}, 270401 (2002).

\bibitem{Lin213901} C. M. Bender, D. C. Brody, and H. F. Jones, Complex Extension of Quantum Mechanics, Phys. Rev. Lett. {\bf 106}, 213901 (2011).
\bibitem{Weimann433} S. Weimann, M. Kremer, Y. Plotnik, Y. Lumer, S. Nolte, K. G. Makris, M. Segev, M. C. Rechtsman, and A. Szameit, Topologically protected bound states in photonic parity-time-symmetric crystals, Nat. Mater. {\bf 16}, 433 (2017).
\bibitem{Rivolta023864} N. X. A. Rivolta, H. Benisty and B. Maes, Topological edge modes with $\mathcal{PT}$ symmetry in a quasiperiodic structure, Phys. Rev. A {\bf 96}, 023864 (2017).

\end{thebibliography}
\end{document}